\pdfoutput=1
\documentclass[a4paper,11pt]{article}
\usepackage{pos}

\providecommand{\repositoryInformationSetup}{} 
\repositoryInformationSetup

\usepackage{graphicx}
\usepackage[T1]{fontenc}
\usepackage[utf8]{inputenc}
\usepackage[UKenglish]{babel}
\usepackage{placeins}
\usepackage[shortlabels]{enumitem}
\usepackage[binary-units=true]{siunitx}
\usepackage{color}
\usepackage{amsmath}
\usepackage{amsthm}
\usepackage{mathtools}
\usepackage{bbold}
\usepackage{braket}
\usepackage[boxruled,lined,commentsnumbered]{algorithm2e}
\usepackage{cleveref}
\usepackage[final]{changes}

\graphicspath{{figures/},{inkscape/},{benchmark/}}

\crefformat{footnote}{#2\footnotemark[#1]#3}

\DeclareMathOperator{\im}{i}
\DeclareMathOperator{\real}{Re}

\DeclareMathOperator{\ord}{\mathcal{O}}

\newcommand{\eto}[1]{\ensuremath{\mathrm{e}^{#1}}}

\newcommand{\ordnung}[1]{\ensuremath{\ord\left(#1\right)}}

\title{Simple Ways to improve Discrete Time Evolution}

\author*[a]{Johann Ostmeyer}

\affiliation[a]{
	Department of Mathematical Sciences,
	University of Liverpool, United Kingdom
}

\emailAdd{J.Ostmeyer@liverpool.ac.uk}

\abstract{
	Suzuki-Trotter decompositions of exponential operators like $\exp(Ht)$ are required in almost every branch of numerical physics. Often the exponent under consideration has to be split into more than two operators, for instance as local gates on quantum computers.
	In this work, we demonstrate how highly optimised schemes originally derived for exactly two operators can be applied to such generic Suzuki-Trotter decompositions.
	After this first trick, we explain what makes an efficient decomposition and how to choose from the large variety available.
	Furthermore we demonstrate that many problems for which a Suzuki-Trotter decomposition might appear to be the canonical ansatz, are better approached with different methods like Taylor or Chebyshev expansions.
	In particular, we derive an efficient and numerically stable method to implement truncated polynomial expansions based on a linear factorisation using their complex zeros.
}

\FullConference{%
	The 40th International Symposium on Lattice Field Theory,\\
	31st July - 4th August, 2022,\\
	Fermilab, Batavia, Illinois, USA
}

\begin{document}

\maketitle


\section{Introduction}

\textbf{(Suzuki-)Trotter decomposition schemes}, or \textbf{Trotterizations}, also referred to as \textbf{splitting methods}, are approximations to operator exponentials of the form
\begin{align}
	\eto{(A+B)h + \ordnung{h^{n+1}}} &= \eto{Aa_1h}\eto{Bb_1h}\eto{Aa_2h}\cdots\eto{Bb_qh}\eto{Aa_{q+1}h}\,,\label{eq:2-stage-decomposition}
\end{align}
where $A,B$ are some non-commuting operators, $h$ is a (small) time step and the $a_i,b_i$ denote some coefficients guaranteeing correctness to order $n$.

Deriving precise Trotter decompositions is not only a challenging mathematical problem, their applications are practically ubiquitous. Just to give a few examples from numerical physics, Trotterization is needed from condensed matter to high energy physics, allowing real and imaginary time evolution~\cite{PRXQuantum.2.010342}, it is a corner stone in hybrid Monte Carlo~\cite{Duane:1987de}, tensor networks~\cite{my_tensor_networks} as well as quantum computing~\cite{adaptive_trotter}.

We remark that sometimes ``symplectic integrator'' is used interchangeably with Trotterization, but this can be very misleading. For more details on the distinctive properties of symplectic integrators see \Cref{sec:deriving_schemes}.

This proceeding expands on the results in our previous work~\cite{trotter_omelyan} where all the technical background can be found that has been omitted here. In the next \Cref{sec:2-to-any-stages} we will reiterate the highly useful method to construct decompositions featuring arbitrarily many operators starting from a two-operator decomposition as in equation~\eqref{eq:2-stage-decomposition}. Closely following Ref.~\cite{OMELYAN2003272}, we will introduce a notion of efficiency for Trotterizations in \Cref{sec:deriving_schemes}. \Cref{sec:taylor_expansion} contains the major novelty of this work, namely an efficient way to implement truncated polynomial (non-Trotter-type) approximations of the matrix exponential that is numerically stable. The approach is based on a factorisation of the polynomial using its zeros in the complex plane. The various methods will be benchmarked in \Cref{sec:numerical_experiments} and a summary of the ``simple ways to improve discrete time evolution'' will be provided in the concluding \Cref{sec:conclusion}.

\section{Adapting 2-stage decompositions to an arbitrary number of stages}\label{sec:2-to-any-stages}
\begin{figure}[bth]
	\centering
	\mbox{\qquad\quad\;\;\,}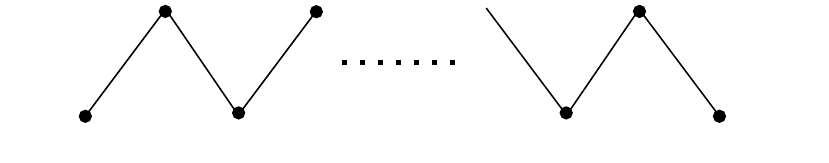\\~\\
	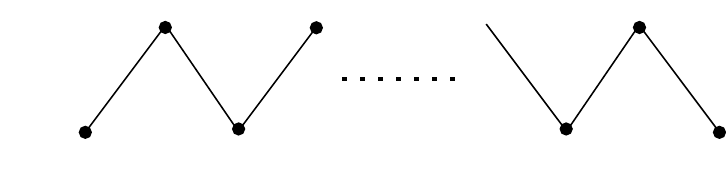
	\caption{Visualisation of the transformation allowing to rewrite a 2-stage decomposition for arbitrarily many stages. In this depiction every dot corresponds to a single operator evaluation or \textit{stage}, a straight line corresponds to a \textit{ramp} and two adjacent ramps form a \textit{cycle}. We start in the top row with a 2-stage decomposition as in eq.~\eqref{eq:2-stage-decomposition} and transform the coefficients $a_i$, $b_i$ into the equivalent $c_i$, $d_i$ as in eq.~\eqref{eq:any-stage-decomposition} using eqs.~\eqref{eq:c1_d1} through~\eqref{eq:cq_dq}. Every ramp can now be populated with more stages in the last row without changing the validity of the decomposition.}\label{fig:stages-visualied}
\end{figure}

Decompositions with two stages, i.e.\ with a Hamiltonian $H=A+B$ consisting of exactly 2 operators as in equation~\eqref{eq:2-stage-decomposition},
have been studied and optimised extensively in literature (e.g.~\cite{Omelyan_2002,BLANES2002313} and many more). However, this is only a special case. Often a Hamiltonian consists of three or more non-commuting operators (an obvious example is the Heisenberg model) and in many cases it is necessary to split it into $\ordnung{L}$ local contributions, e.g.\ to apply gates to tensor network or quantum simulations. In these cases only less optimised schemes like those by Hatano and Suzuki~\cite{Hatano_2005} or Yoshida~\cite{YOSHIDA1990262} have been applicable so far. Now recently, we have shown that \textit{every scheme of order $n$ applicable to two stages defines an order $n$ scheme for an arbitrary number of stages}~\cite{trotter_omelyan}. This allows to adapt the valuable work done on 2-stage decompositions so far and it furthermore provides a simple means to derive efficient new methods in the future since it is sufficient to analyse the method's properties in the simple case of two stages.
\added{The idea behind the transformation required for this adaptation has been visualised in figure~\ref{fig:stages-visualied}.}

More rigorously, the theorem proved in~\cite{trotter_omelyan} reads as follows.
Let the coefficients $a_i$ and $b_i$ as in equation~\eqref{eq:2-stage-decomposition} define a 2-stage decomposition of order $n$. Then the decomposition
\begin{align}
	\eto{h\sum\limits_{k=1}^{\Lambda}A_k + \ordnung{h^{n+1}}} &= \left(\prod_{k=1}^{\Lambda}\eto{A_kc_1h}\right) \left(\prod_{k=\Lambda}^{1}\eto{A_kd_1h}\right) \cdots \left(\prod_{k=1}^{\Lambda}\eto{A_kc_qh}\right) \left(\prod_{k=\Lambda}^{1}\eto{A_kd_qh}\right)\label{eq:any-stage-decomposition}
\end{align}
of $\Lambda$ non-commuting operators $A_k$ into ordered products $\prod_{k=1}^{\Lambda}X_k\equiv X_1X_2\cdots X_\Lambda$ is exact to order $n$ as well and the coefficients $c_i$, $d_i$ are given by
\begin{align}
	c_1 &= a_1\,, & d_1 &= b_1 - c_1\,,\label{eq:c1_d1}\\
	c_2 &= a_2 - d_1\,,& d_2 &= b_2 - c_2\,,\\
	&\;\;\vdots & &\;\;\vdots\nonumber\\
	c_q &= a_q - d_{q-1}\,,& d_q &= b_q - c_q\,.\label{eq:cq_dq}
\end{align}


\section{Analysing the efficiency of decomposition schemes}\label{sec:deriving_schemes}
For the rest of this work we will consider symmetric decomposition schemes only. Besides their often useful reversibility, symmetric schemes are simply more efficient for evolutions over long times $t$, i.e.\ when $t/h\gg1$ which is the physically interesting and computationally challenging case. The reason is that any non-symmetric and therefore odd order $n-1$ scheme can be elevated to even order $n$ without any drawbacks by reverting the sequence of coefficients $a_i$, $b_i$ (or $c_i$, $d_i$) in every second step. The established symmetry guarantees an even order $n$. In case the integer $t/h$ was an even number to start with, the computational effort remains the same. If $t/h$ was odd, it has to be incremented by one at the relative additional computational cost of $h/t\ll1$. In both cases the induced symmetric even order $n$ decomposition is strictly more efficient than the initial non-symmetric one.

Omelyan, Mryglod and Folk~\cite{Omelyan_2002} as well as Blanes and Moan~\cite{BLANES2002313} derived a number of highly optimised schemes. Closely following the approach by Omelyan et al.\ we will now define what requirements exactly a decomposition scheme has to fulfil to be of order $n$ and what it means for such a scheme to be efficient.


Following the derivations in Refs.~\cite{Omelyan_2002,OMELYAN2003272}, we can write down a basis for the leading order errors of a decomposition
\begin{align}
	\eto{(A+B)h + \ord_1h + \ord_3 h^3 + \ord_5 h^5 + \cdots} &= \eto{Aa_1h}\eto{Bb_1h}\cdots\eto{Bb_qh}\eto{Aa_{q+1}h}
\end{align}
in terms of commutators
\begin{align}
	\ord_1 &= (\nu-1) A + (\sigma-1)B\,,\label{eq:ord_1_errors}\\
	\ord_3 &=  \alpha [A, [A, B]] + \beta [B, [A, B]]\,,\label{eq:ord_3_errors}\\
	\begin{split}
		\ord_5 &= \gamma_1 [A, [A, [A, [A, B]]]] + \gamma_2 [A, [A, [B, [A, B]]]] + \gamma_3 [B, [A, [A, [A, B]]]]\\
		&\quad + \gamma_4 [B, [B, [B, [A, B]]]] + \gamma_5 [B, [B, [A, [A, B]]]] + \gamma_6 [A, [B, [B, [A, B]]]]\,.
	\end{split}\label{eq:ord_5_errors}
\end{align}

A decomposition is valid if $\nu=\sigma=1$, that is $\ord_1=0$. This can easily be guaranteed by the intuitive condition $\sum_i a_i=\sum_i b_i=1$ (or equivalently $\sum_i(c_i+d_i)=1$). In order to construct a 4th order decomposition, $\alpha=\beta=0$ has to be satisfied additionally. For a 6th order scheme all the $\gamma_j$ have to vanish, and so on.

\added{We remark that symplectic integrators (or Runge–Kutta–Nyström methods) deal with the special case of $[B, [B, [B, A]]] = 0$ leading to significantly less contributions to all the operators $\ord_k$ with $k\ge 5$ than in the general case discussed here. This allows for more efficient but specialised methods, in particular higher order symplectic decompositions $n\ge 6$ require less cycles $q$ than general Trotterizations. This distinction has been discussed in detail by Blanes and Moan~\cite{BLANES2002313}. A comprehensive list of efficient symplectic integrators up to order $n\le 6$ has been provided by Omelyan~\cite{OMELYAN2003272}, we especially recommend Table~2 therein.}

A decomposition is efficient if its leading order errors are small compared to the number $q$ of cycles\footnote{$q$ is also often called `stages' or `number of force calculations' in the literature.} it requires. We assume all the commutator products in the equations~(\ref{eq:ord_1_errors}-\ref{eq:ord_5_errors}) to be mutually orthogonal so that the euclidean norm over the spanned vector space estimates the total errors. Note that orthogonality is not guaranteed, but for high enough dimensional operators the overlap will be negligible in general. In particular we define the efficiencies following Omelyan~\cite{OMELYAN2003272}
\begin{align}
	\mathrm{Eff}_2 &= \frac{1}{q^2\sqrt{|\alpha|^2+|\beta|^2}}\,,&
	\mathrm{Eff}_4 &= \frac{1}{q^4\sqrt{\sum_{j=1}^6|\gamma_j|^2}}\,.\label{eq:eff-formula}
\end{align}
\added{Higher order efficiencies $\mathrm{Eff}_n$ are defined analogously via the leading order error and the respective number of cycles.}
\deleted{It does not make sense to directly compare efficiencies of different decompositions across orders. The optimal order to choose depends foremost on the desired accuracy.}
Within an order the efficiency is directly proportional to the precision obtained at any given computational effort and a higher efficiency is generally better. More details on comparisons between orders and a detailed list of Suzuki-Trotter decompositions can be found in Ref.~\cite{trotter_omelyan}.

\section{Truncated polynomial expansion}\label{sec:taylor_expansion}
We now present a totally different, not strictly unitary, approach based on the truncated Taylor expansion
\begin{align}
	\eto{H h} &= \lim_{k\rightarrow\infty}\sum_{i=0}^k\frac{\left(H h\right)^i}{i!}\,,\label{eq:taylor_expansion}
\end{align}
or Chebyshev expansion
\begin{align}
	\eto{H h} &= \lim_{k\rightarrow\infty}\sum_{i=0}^k \mu_i T_i\left(\frac{H}{\Gamma}\right)\,,\quad \mu_0 = I_0(\Gamma h)\,,\; \mu_i = 2I_i(\Gamma h)\,,\label{eq:chebyshev_expansion}
\end{align}
where the $T_i$ denote the Chebyshev polynomials, $I_i$ the modified Bessel functions of first kind and $\Gamma>0$ large enough (as defined below).
Note that the coefficients $\mu_i$ in equation~\eqref{eq:chebyshev_expansion} are different for real- and imaginary time evolution (use $I_i(\im \Gamma h) = \im^{-i} J_i(\Gamma h)$ with the (unmodified) Bessel functions $J_i$ in the former case).

The cutoff $k$ has to be chosen so that all eigenvalues $\lambda$ of $H$ are sufficiently suppressed
\begin{align}
	\left|\frac{\left(\lambda_\text{max}(H) h\right)^{k}}{(k+1)!}\right| &< \varepsilon & \text{(Taylor)}\,,\label{eq:prec_req}\\
	\left|\mu_{k+1}\right| &< \varepsilon & \text{(Chebyshev)}\,,
\end{align}
where $\varepsilon$ is the desired relative precision. Technically a cutoff $k$ large enough guarantees super-exponential convergence independently of the step size $h$ for both, Taylor and Chabyshev expansions. In practice, however, long time evolutions have to be subdivided into factors of limited step size $h$. In particular, the Taylor addends~\eqref{eq:taylor_expansion} can span many orders of magnitude so that finite precision arithmetics leads to significant errors.
In Ref.~\cite{trotter_omelyan} we presented more details on the derivation of the optimal step size $h$ and cutoff $k$ assuming the direct summation of the terms in the Taylor series~\eqref{eq:taylor_expansion}.

\begin{figure}
	\centering
	\includegraphics[width=.48\textwidth]{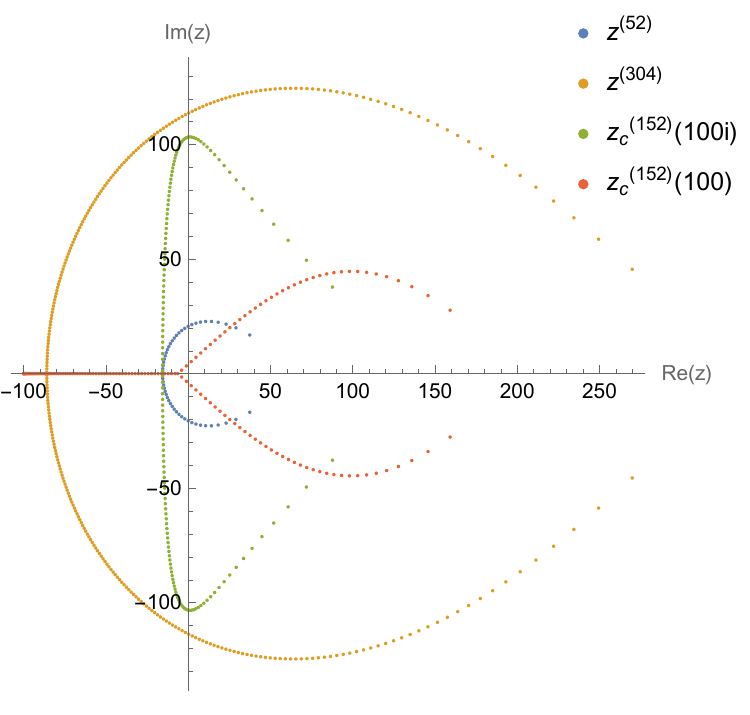}%
	\hfill%
	\includegraphics[width=.48\textwidth]{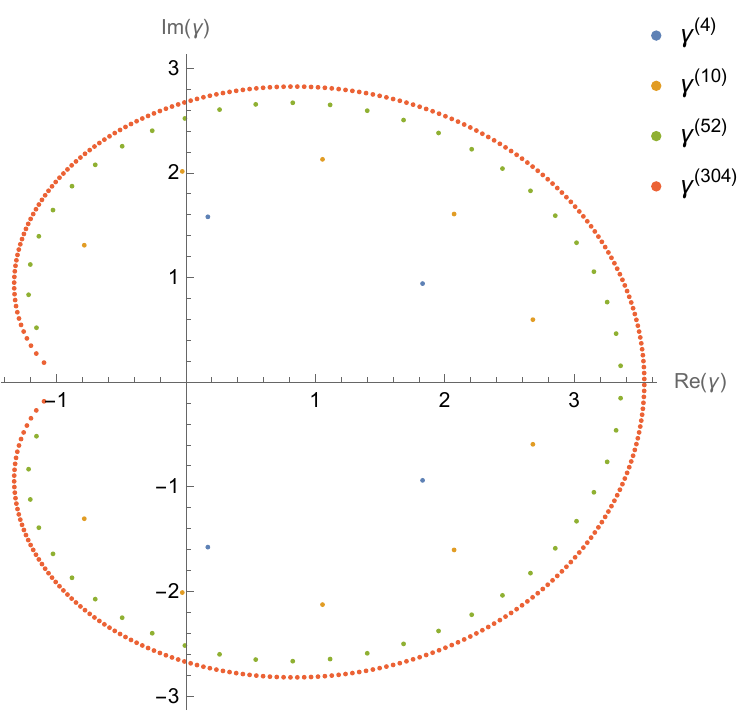}
	\caption{Distribution of Taylor and Chebyshev expansion coefficients for the exponential function in the complex plane. A factorisation~\eqref{eq:exp_factorisation} is exact up to double machine precision within $|z|\le10$ when using $z^{(52)}$, within $|z|\le100$ using $z^{(304)}$, in the interval $z\in[-100\im,100\im]$ using $z_c^{(152)}(100\im)$, and in $z\in[-100,100]$ using $z_c^{(152)}(100)$. Left: Zeros $z^{(k)}$ of the truncated Taylor series~\eqref{eq:taylor_expansion} and zeros $z_c^{(k)}(\Gamma h)$ of the Chebyshev polynomial~\eqref{eq:chebyshev_expansion} best describing the interval $[-\Gamma h,\Gamma h]$ truncated at $k$-th order, respectively. Right: Coefficients $\gamma^{(k)}=-k/z^{(k)}$ in the factorisation of the Taylor series.}\label{fig:zeros-coeffs}
\end{figure}

Here, we put forward a fundamentally different method for the evaluation of the truncated series. Since we are always dealing with finite order $k$ polynomials, they have $k$ zeros $z_i^{(k)}$ in the complex plane and the exponential can be approximated as
\begin{align}
	\eto{Hh} &= \prod_{i=1}^{k}\left(1+\frac{\gamma_i^{(k)}}{k}Hh\right) + \ordnung{h^{k+1}}\label{eq:exp_factorisation}\\
	&\propto \prod_{i=1}^{k}\left(Hh-z_i^{(k)}\right) + \ordnung{h^{k+1}}\,,\\
	\gamma_i^{(k)} &\equiv -\frac{k}{z_i^{(k)}} \,.
\end{align}
In particular the zeros of the truncated Taylor series are very well behaved falling on the Szeg\H{o} curve~\cite{VARGA2010298}
\begin{align}
\lim_{k\rightarrow\infty} \frac 1k z^{(k)} = \left\{z: |z\eto{1-z}|=1,|z|\le1\right\}\,.
\end{align}
Both, the zeros (left) and their corresponding coefficients (right) have been visualised in figure~\ref{fig:zeros-coeffs}. The distribution of the zeros intuitively forms a boundary for the region in which the exponential function is well approximated by the respective truncation since $\exp(z)$ itself is never zero and a zero of the polynomial marks a breakdown of the approximation. Clearly, the Taylor series provides a universal approximation in the complex plane within a given radius (asymptotically $k/e$) while the Chebyshev polynomials only yield a good approximation on either the real or the imaginary axis. However, on this one axis the Chebyshev polynomials can approximate an interval that is asymptotically as large as $[-k,k]$ which makes them a factor $e\approx\num{2.7}$ more efficient than the Taylor expansion in the limit of large $k$.

In practice, an implementation using the factorisation~\eqref{eq:exp_factorisation} is always preferable to a direct summation~\eqref{eq:taylor_expansion} because it does not suffer the aforementioned numerical instability. Less obviously, the factorised version is also more memory efficient than either Taylor or Chebyshev summation. Furthermore it has a reduced runtime compared to the stable iterative construction
\begin{align}
	T_0(H) &= \mathbb{1}\,,\\
	T_1(H) &= H\,,\\
	T_{i+1}(H) &= 2H\cdot T_i(H) - T_{i-1}(H)
\end{align}
of Chebyshev polynomials. For high accuracy, however, it if of crucial importance to choose a good order for the coefficients $\gamma_i^{(k)}$, even though theoretically all the factors in the product~\eqref{eq:exp_factorisation} commute. We find empirically that best results are achieved when coefficients are always grouped together in tuples, so that the sum over the coefficients within each tuple is roughly the same. In this manner, the $\gamma_i^{(k)}$ can be thought of as individual sub-steps that should be arranged in a way maintaining a roughly constant progress.

Since the coefficients of the factorised polynomials are real, all the zeros $z_i^{(k)}$ and coefficients $\gamma_i^{(k)}$ come in complex conjugate pairs. These pairs play a special role in the aforementioned ordering of coefficients. They should never be separated. In fact, numerical accuracy is increased when these pairs are jointly evaluated in a quadratic term with real coefficients
\begin{align}
	\left(1+\frac{\gamma_i^{(k)}}{k}Hh\right)\left(1+\frac{\bar\gamma_i^{(k)}}{k}Hh\right) &= 1+2\frac{\real{\gamma_i^{(k)}}}{k}Hh + \frac{\left|\gamma_i^{(k)}\right|^2}{k^2}\left(Hh\right)^2\,.
\end{align}
In most classical implementations this quadratic version will be favourable over the fully factorised one, especially if the complex numbers can be completely avoided this way. The only drawback is the increased memory footprint that can become relevant in some cases, especially if the squared hamiltonian $H^2$ has to be constructed explicitly.

Note that the number $k$ is not equivalent with the number of cycles. This relation strongly depends on the implementation. For instance in the case we are discussing below (see \Cref{sec:numerical_experiments}) the implementation with cutoff $k$ has approximately the runtime of $q\approx k/6$ cycles.


\section{Numerical experiments}\label{sec:numerical_experiments}
\begin{figure}[ht]
	\centering
	\resizebox{0.98\textwidth}{!}{{\large\input{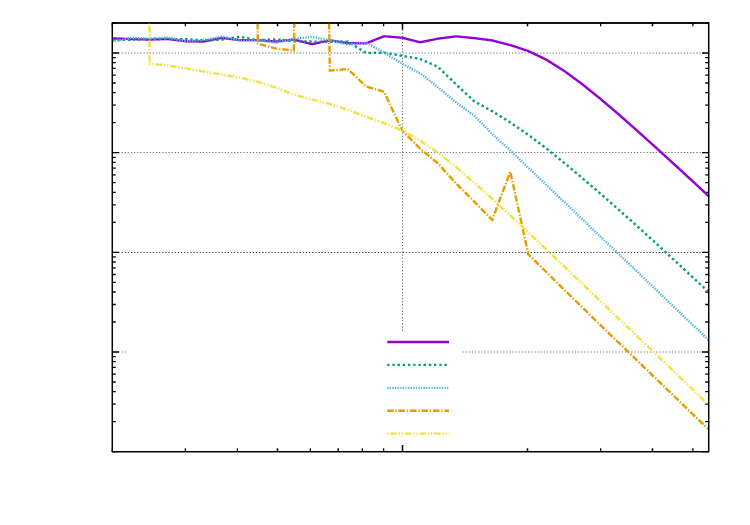}\input{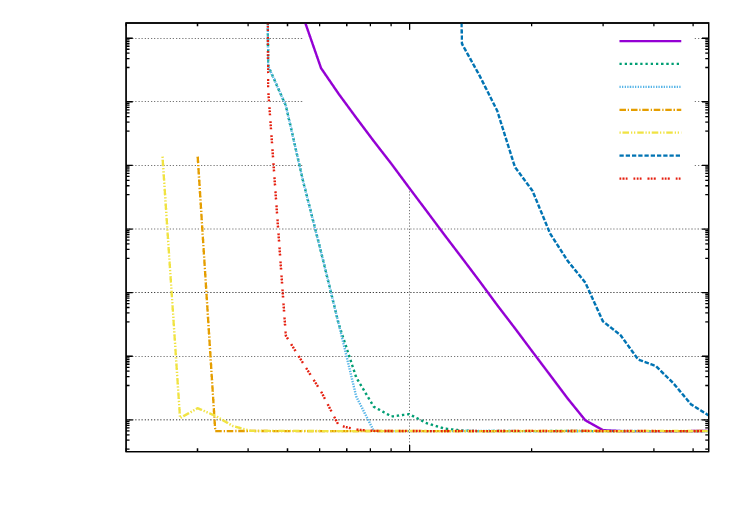}}}
	\caption{Collection of order $n=4$ schemes (left) and truncated polynomial expansions (right). Error, i.e.\ the difference between the exact time evolution operator and the respective decomposition in Frobenius norm, as a function of computational cost, i.e.\ number of cycles $q$ (assuming $q=k/6$ for non-Trotter type methods) divided by the time step $h$. Following the name of the scheme we write the order / cycles tuple $(n,q)$ or the cutoff / method tuple $(k,\cdot)$, respectively. Here the method can be either `$\sum$' as in eqs.~\eqref{eq:taylor_expansion} and~\eqref{eq:chebyshev_expansion} or `$\prod$' as in eq.~\eqref{eq:exp_factorisation}}\label{fig:err_const_t}
\end{figure}

We use the Heisenberg XXZ-model~\cite{real_time_dos} as a test for the decomposition schemes. Details on the model and the error estimation can be found in~\cite{trotter_omelyan}. In short, we compute the exact time evolution operator $\eto{-\im H t}$ using exact diagonalisation and calculate its difference to the respective approximation in Frobenius norm. The results are visualised in figure~\ref{fig:err_const_t}.
All the errors have been obtained after the same evolution time of $t=100$ and the axis labelled $q/h$ is equivalent to the de-facto computational costs. For the Trotter-type schemes the minimal number of 3 stages has been used.

The fourth order schemes benchmarked here (fig.~\ref{fig:err_const_t}, left) have been catalogued in~\cite{trotter_omelyan} including their theoretical efficiencies~\eqref{eq:eff-formula}. In accordance with said efficiencies, they form a clear hierarchy that is mirrored by our numerical experiments. The decomposition by Forest and Ruth~\cite{FOREST1990105} is clearly worst by a margin. We quote it here because it is still popular in the literature for its simplicity, but we emphasise that it should never be used. Next comes Suzuki's standard scheme~\cite{Hatano_2005} which can be considered a baseline to which every other decomposition has to be compared. Applying our transformation from equations~\eqref{eq:c1_d1} through~\eqref{eq:cq_dq} to the highly optimised scheme by Blanes and Moan~\cite{BLANES2002313}, allows to significantly outperform Suzuki's decomposition. This makes Blanes and Moan's scheme the most efficient strictly unitary fourth order decomposition to date. Our non-unitary method~\cite{trotter_omelyan} employing complex-valued coefficients as well as a fourth order Taylor expansion both easily outperform the unitary methods. It is interesting to observe that Taylor's method gives relatively good results even for extremely low computations cost, even though it is asymptotically inferior to the non-unitary Trotter-type decomposition.

We remark that this reasonably good performance at low cost is a characteristic feature of low order Taylor series setting them apart from Chebyshev expansions. In this regime a Chebyshev expansion is uniformly mediocre, while the Taylor series still describes most of the eigenvalues of $H$ very well because they tend to be clustered around zero. This makes Taylor the ideal choice when the target precision is not too high, Chebyshev on the other hand is preferable in the regime of high accuracy and high orders.

The same conclusion is supported in the right panel of figure~\ref{fig:err_const_t} where results from different truncated polynomials are summarised. As expected, we find that the efficiency of the Taylor series increases with higher orders / cutoffs $k$ and that they in turn yield to Chebyshev polynomials in terms of performance. It is absolutely essential for this trend that the factorised version `$\prod$' (eq.~\eqref{eq:exp_factorisation}) is used rather than the sum `$\sum$' (eqs.~\eqref{eq:taylor_expansion}, \eqref{eq:chebyshev_expansion}). While in the case of Chebyshev the factorisation merely reduces computational costs, for a Taylor series it can mean the difference between a correct result and utter nonsense. In double precision, the difference is bound to occur for every cutoff $k>17$, it is clearly visible for $k=52$, and it becomes a dominant feature for $k=304$.

On a side note, the identical error plateau reached by all Taylor and Chebyshev methods strongly indicates that the limiting factor in this case is no longer the respective quoted method but the exact diagonalisation which is also performed to finite precision.

\section{Conclusion}\label{sec:conclusion}
In this work we have presented a variety of methods for discrete time evolution. The focus lies on representative fourth order Trotterizations and truncated polynomial expansions. It is not the aim of this work to provide a comprehensive overview over different Trotter decomposition and the interested reader is referred to our previous work~\cite{trotter_omelyan} for this purpose as well as for a flow chart providing a simple guideline to choose a close to optimal method in a given scenario.

The main novelty setting this work apart from~\cite{trotter_omelyan} is the introduction and discussion of factorised formulae~\eqref{eq:exp_factorisation} for the efficient and numerically stable implementation of truncated polynomial series in \Cref{sec:taylor_expansion}. We showed how, based on the complex zeros of the polynomial, a product featuring only coefficients of order one can be constructed and demonstrated that this construction is always preferable to the naive summation of the different order contributions.

In summary, efficient discrete time evolution can be boosted rather easily if the following points are considered prior to deciding on an implementation: Use a Chebyshev or Taylor expansion (in this order) whenever possible and implement it in the factorised form~\eqref{eq:exp_factorisation}. Do not be afraid to use complex-valued coefficients in Trotter decompositions as they can significantly improve precision at the price of marginally violating unitarity. Adapt highly optimised 2-operator decompositions like that by Blanes and Moan~\cite{BLANES2002313} or our non-unitary schemes for any required number of operators using the transformation detailed in equations~\eqref{eq:c1_d1} through~\eqref{eq:cq_dq} and visualised in figure~\ref{fig:stages-visualied}.

Finally, we remark that the theoretical error prediction for a given time evolution method is still anything but accurate and subject to ongoing research. So long, the only way to guarantee that the optimal solution to a given problem has been chosen, remains numerical testing of that particular problem.


\section*{Acknowledgements}
This work was funded in part by the STFC Consolidated Grant ST/T000988/1.
The author thanks Pavel Buividovich, Michael Lubasch, Tom Luu, and David Martí-Pete for their helpful comments.

The entire code (implemented in \texttt{C} with a front-end in \texttt{R}) and data required to reproduce the results and plots of this paper have been published under open access and can be found in~\cite{johann_ostmeyer_2022_7268893}. The coefficients $\gamma^{(k)}$ have been computed in  \texttt{Mathematica} and a notebook allowing to calculate these coefficients to arbitrary precision is included in the code.

\FloatBarrier
\bibliographystyle{JHEP}
\bibliography{bibliography}

\end{document}